\documentclass[runningheads]{llncs}
\usepackage{graphicx}
\usepackage{textcomp}
\usepackage{comment}

\usepackage{color}
\usepackage{graphicx}
\usepackage{amsmath}
\usepackage{amssymb}
\usepackage{booktabs}
\usepackage{verbatim}
\usepackage{import}
\usepackage{multirow}
\usepackage{hyperref}
\usepackage[capitalize]{cleveref}
\usepackage{caption}
\usepackage{subcaption}
\usepackage[utf8]{inputenc}
\usepackage[english]{babel}
\usepackage{csquotes}
\usepackage{adjustbox}
\usepackage{float}
\restylefloat{table}
\usepackage{placeins}
\usepackage{xspace}
\makeatletter
\DeclareRobustCommand\onedot{\futurelet\@let@token\@onedot}
\def\@onedot{\ifx\@let@token.\else.\null\fi\xspace}

\makeatother

\usepackage{todonotes}

\usepackage{float}
\floatstyle{plaintop}
\restylefloat{table}
\usepackage{mdwlist}
\usepackage{array}

\begin{document}
\titlerunning{An investigation into the causes of race bias}

\title{An investigation into the causes of race bias in AI-based cine CMR segmentation}

\newcommand*\samethanks[1][\value{footnote}]{\footnotemark[#1]}
\newcommand{\rowstyle}[1]{\gdef\currentrowstyle{#1}%
  #1\ignorespaces
}

\author{Tiarna Lee \inst{1}
\and Esther Puyol-Ant\'on \inst{1, 4} \and 
Bram Ruijsink \inst{1,2}
\and Sebastien Roujol \inst{1}
\and Theodore Barfoot \inst{1}
\and Shaheim Ogbomo-Harmitt \inst{1}
\and Miaojing Shi \inst{3}
\and Andrew P. King \inst{1} }
\authorrunning{T Lee et al.}   
\institute{School of Biomedical Engineering \& Imaging Sciences, King\textquotesingle s College London, UK. \and Guy’s and St Thomas’ Hospital, London, UK. \and College of Electronic and Information Engineering, Tongji University, China. \and HeartFlow Inc, London, UK.}
\maketitle              

\begin{abstract} 
Artificial intelligence (AI) methods are being used increasingly for the automated segmentation of cine cardiac magnetic resonance (CMR) imaging. However, these methods have been shown to be subject to race bias, i.e. they exhibit different levels of performance for different races depending on the (im)balance of the data used to train the AI model. In this paper we investigate the source of this bias, seeking to understand its root cause(s) so that it can be effectively mitigated. We perform a series of classification and segmentation experiments on short-axis cine CMR images acquired from Black and White subjects from the UK Biobank and apply AI interpretability methods to understand the results. In the classification experiments, we found that race can be predicted with high accuracy from the images alone, but less accurately from ground truth segmentations, suggesting that the distributional shift between races, which is often the cause of AI bias, is mostly image-based rather than segmentation-based. The interpretability methods showed that most attention in the classification models was focused on non-heart regions, such as subcutaneous fat. Cropping the images tightly around the heart reduced classification accuracy to around chance level. Similarly, race can be predicted from the latent representations of a biased segmentation model, suggesting that race information is encoded in the model. Cropping images tightly around the heart reduced but did not eliminate segmentation bias. We also investigate the influence of possible confounders on the bias observed.

\end{abstract}

\section{Introduction}
\label{sec:intro}

Cardiac Magnetic Resonance (CMR) imaging is widely used to acquire images for diagnosis and prognosis of cardiovascular conditions. Artificial intelligence (AI) methods are increasingly being used to automate the estimation of functional biomarkers from cine CMR by automatic delineation (segmentation) of cardiac structures \cite{Mariscal-Harana2023AnDatabases}, \cite{Davies2022PrecisionLearning}, \cite{Ruijsink2020FullyFunction}. However, recent work has shown that AI CMR segmentation models can exhibit different levels of performance for different protected groups, such as those based on race \cite{Puyol-Anton2021FairnessSegmentation}, \cite{Puyol-Anton2022FairnessSegmentation}, \cite{Lee2022ASegmentation} or sex \cite{Lee2023AnSegmentation} (i.e. they can be biased). In order to properly address this bias, it is important to understand its causes, but these are not yet well understood. This paper presents an investigation into the causes of race bias in AI-based CMR segmentation.

\section{Contributions}
The contribution of this work is to investigate the cause of bias in AI-based CMR segmentation models. We show that the main source of bias is in the image content outside of the heart region and that bias can be reduced by cropping the images before training the AI models.

\section{Methods}
\subsection{Dataset}
\label{sec:materials}

The dataset used in the experiments described in this paper comprised cine short axis (SAX) CMR images from 436 subjects from the UK Biobank \cite{Peterson2016}. For each subject, typically 7 – 13 SAX slices were available at 50 time frames covering the cardiac cycle. The demographic information of the subjects can be found in \cref{tab: EHJDH subject characteristics}.

\begin{table}
\centering

\caption{Clinical characteristics of subjects used in the experiments. Mean values are presented for each characteristic with standard deviations given in brackets. Statistically significant differences between subject groups and the overall average are indicated with an asterisk * (p $<$ 0.05) and were determined using a two-tailed Student’s t-test.}
\begin{adjustbox}{width=0.8\columnwidth,center}
\begin{tabular}{|c|c|c|c|}
\hline
\textbf{Health measure}       & \textbf{Overall}     & \textbf{White}       & \textbf{Black}                            \\
\hline \# subjects          & 436         & 218         & 218                              \\
\hline Age (years           & 58.9 (7.0)  & 58.9 (7.0)  & 58.8 (6.9)                       \\
\hline Standing height (cm) & 80.6 (16.6) & 79.3 (17.0) & 82.0 (16.1)                      \\
\hline Body Mass Index      & 27.7 (4.9)  & 26.9 (4.6)* & 28.6 (5.1)*  \\ \hline
\end{tabular}
\label{tab: EHJDH subject characteristics}
\end{adjustbox}
\end{table}

The subjects selected for potential inclusion from the full UK Biobank CMR cohort were those with available manual ground truth segmentations (4928 out of 78,166 subjects) \cite{Petersen2016ReferenceCohort}. The manual segmentations were of the left ventricular blood pool (LVBP), left ventricular myocardium (LVM), and right ventricular blood pool (RVBP) and were performed for the end diastole (ED) and end systole (ES) images. Therefore, only the ED and ES frames were used in our experiments. Manual segmentation was performed by outlining the LV endocardial and epicardial borders and the RV endocardial border using cvi42 (version 5.1.1, Circle Cardiovascular Imaging Inc., Calgary, Alberta, Canada). A panel of ten experts was provided with the same guidelines and one expert annotated each image. The selection of images for annotation included subjects with different sexes and races and was randomised. The experts were not provided with demographic information about the subjects.

From the available data, a cohort of 218 Black subjects was selected for use in all experiments. This cohort was chosen to have 109 males (all available Black males) and 109 females to minimise the impact of possible sex bias. To select the White subjects a matched pairs design was used, in which White subjects with matching age (± 1 year) and sex to each Black subject were chosen at random from the available pool of 4690 White subjects with ground truth segmentations. For each subject, the ED and ES frames of all SAX CMR slices and their corresponding ground truth segmentations were utilised in the experiments. Demographic and health data from the subjects was acquired from the UK Biobank database including the subjects’ age, standing height, weight, body mass index (BMI), resting heart rate, systolic and diastolic blood pressure, left ventricular stroke volume (LVSV), left ventricular ejection fraction (LVEF), left ventricular end diastolic mass (LVEDM), HDL cholesterol, cholesterol, diabetes status, hypertension status, hypercholesterolemia status, smoking status and date of the MRI scan. 

\subsection{Models used}

To perform the investigations into the source of bias we employ two types of AI model: a classification model and a segmentation model.

ResNet-18 is a deep convolutional neural network (CNN) for classification consisting of 18 layers \cite{He2016DeepRecognition}. The network has residual blocks and skip connections which can be used to form deep networks. For the classification experiments (Experiments 1 and 2 in the Results), the model was trained for 100 epochs with an initial learning rate of 0.001 which decreased by a factor of 10 every 50 epochs. The loss function used was binary cross entropy and the model was optimised using stochastic gradient descent. The batch size was 16. The images were augmented using random mirroring, rotating, scaling and translation. As the images are greyscale, no colour intensity transformations were used. Each model was trained 10 times with different random seeds and train/validation splits and the mean and standard deviation for these 10 runs is reported.

For the classification network, we also employed the gradient-weighted class activation mapping method, or GradCAM, which is a visualisation and interpretability method \cite{SelvarajuGrad-CAM:Localization}. The gradients of the target class (in our case, race) in the last convolutional layer of the classification network were visualised to produce a heatmap which shows the areas of an image that were most important for the classification decision. 

For the segmentation experiments, we used nnU-Net, a self-adapting framework for segmentation of biomedical images \cite{Isensee2020}. The network automatically adapts to the imaging modality and changes training parameters such as the patch size, batch size and image resampling. The nnU-Net v1 model consists of an encoder and decoder structure which form a ``U'' shape, allowing the network to learn a more abstract representation of the images. For the segmentation experiments (Experiments 1 and 2 in the Results), the model was trained for 500 epochs with an initial learning rate of 0.01. The loss function used was a combined Dice and cross entropy loss. The model was optimised using stochastic gradient descent with a ‘poly’ learning rate schedule, where the initial learning rate was 0.01 and the Nesterov momentum was 0.99. A batch size of 16 was used. During training, data augmentation was applied to the images including mirroring, rotation and scaling. Cross-validation was performed on the training set, resulting in five models, which were used as an ensemble for inference on the test set.

\subsection{Statistical evaluation}
Classification accuracy was evaluated using overall accuracy, sensitivity and specificity. Differences in performances were evaluated using a two-tailed Student’s t-test of the accuracies of the 10 runs. Segmentation performance was evaluated using the Dice Similarity Coefficient (DSC) which measures the overlap between ground truth and predicted segmentations where 1 is a perfect overlap and 0 is no overlap. Confounder analysis was performed using linear regression models in SPSS Statistics (IBM Corp. Released 2023. IBM SPSS Statistics for Macintosh, Version 29.0.2.0 Armonk, NY: IBM Corp).

\section{{Results}}
\label{sec:results} 

The experiments performed using the data and models described above aimed to investigate three aspects of the bias in AI CMR segmentation performance as detailed below.

\subsection{Experiment 1: Source of bias}
Bias in AI models is often the result of a distributional shift in the data of subjects in different protected groups. Combined with imbalance in the training data, these distributional shifts can lead to bias in performance of AI models \cite{Lee2022ASegmentation,Larrazabal2020}. However, the distributional shift can be in the images, the ground truth segmentations or a combination of both. Understanding the origin of the bias in trained segmentation models is important when deciding on strategies to address it. Therefore, the first experiment aimed to assess the extent of the distributional shift between the CMR images and/or the ground truth segmentations.

To quantify the extent of the distributional shifts, we trained ResNet-18 models to classify the race of the subject (White vs Black) from a single SAX CMR image and/or segmentation. The SAX CMR images and ground truth segmentations of the 218 Black and 218 White subjects were randomly split at the subject level into training and test datasets with 176 and 84 subjects respectively (ensuring that both images from each matched pair were in the same split). 

\begin{figure}[t]
\includegraphics[width=12cm]{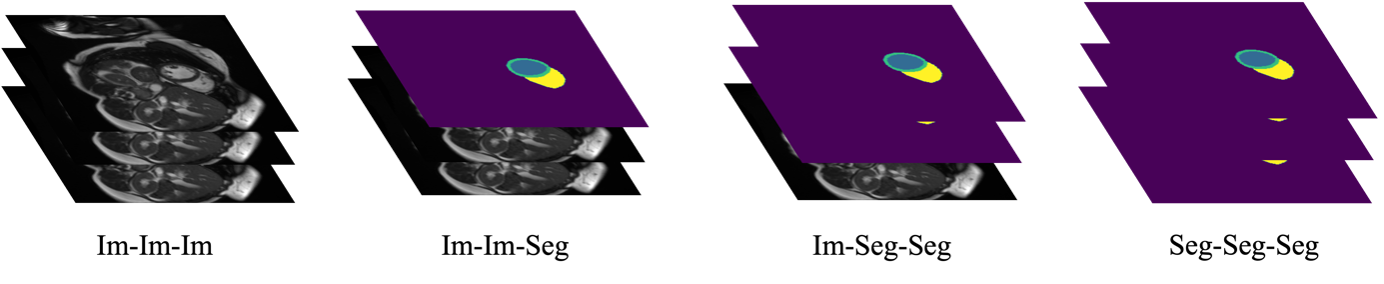}
\centering
\caption{An illustration of the combination of images and segmentations used as input to the protected attribute classifiers}
\label{fig: EHJDH ims diagram}
\end{figure}

The classifier was trained with three channels as input. To assess the relative distributional shifts between images and ground truth segmentations we used four different combinations of images (Im) and segmentations (Seg): Im-Im-Im, Im-Im-Seg, Im-Seg-Seg, and Seg-Seg-Seg, as illustrated in \cref{fig: EHJDH ims diagram}. 

\cref{tab: classification acc} shows the results for classifying Black and White race subjects. The highest accuracies were achieved when images were used, either on their own or in combination with segmentations. The accuracy of the Seg-Seg-Seg dataset was the lowest but still higher than random chance. Using a two-tailed Student’s t-test between the accuracies of the 10 runs for each of the datasets, the only significant differences were between the Seg-Seg-Seg dataset and the other three datasets which contained images (p $<$ 0.0001 for all).

 \begin{table}
\centering
\begin{adjustbox}{width=0.9\columnwidth,center}

\caption{Accuracy for experiment on classifying the subjects by race (Black and White). The results show the mean (standard deviation) over 10 repeat runs. The highest result for each measure is shown in bold.}
\begin{tabular}{|c|c|c|c|}
\hline \textbf{Image type} & \textbf{Accuracy}       & \textbf{Sensitivity}             & \textbf{Specificity}                \\ \hline
Im-Im-Im            & \textbf{0.959 (0.004)~} & \textbf{0.966 (0.013)}~\textbf{} & 0.951 (0.014)~                      \\
\hline Im-Im-Seg           & 0.957 (0.010)~          & 0.959 (0.011)~                   & \textbf{0.956 (0.018)}~\textbf{}    \\
\hline Im-Seg-Seg          & 0.955 (0.007)~          & 0.961 (0.013)~                   & 0.948 (0.010)~                      \\
\hline Seg-Seg-Seg         & 0.742 (0.005)~          & 0.727 (0.011)~                   & 0.765 (0.020)~ \\ \hline
\end{tabular}
\label{tab: classification acc}
\end{adjustbox}
\end{table}

The conclusion of this first experiment is that the majority of the distributional shift between races lies in the images rather than the ground truth segmentations.

The next experiment aimed to assess whether, and the degree to which, this distributional shift is also encoded in trained segmentation models. To answer these questions, we used an approach similar to that described in \cite{Glocker2023AlgorithmicModels}. First, nnU-Net models were trained to segment CMR images using training data with varying levels of race imbalance. Next, the decoder part of the trained networks was removed. The test set was then fed through the encoder part of the network to produce a latent vector for each test subject. We then used principal components analysis (PCA) to reduce the dimensionality of these latent representations and visualised the results. Furthermore, we investigated whether race could be separated in the reduced dimensional space using logistic regression classification.

The quantitative results for this experiment are shown in \cref{tab: logistic regression acc}. The PCA reduced-dimensional representations of the images could be classified with a high accuracy of approximately 95\% for all models. Visual illustrations of the classification of protected attributes can be found in \cref{fig: latent space classification}.

\begin{table}
\centering
\begin{adjustbox}{width=0.7\columnwidth,center}

\caption{Accuracy of a logistic regression model classifying the PCA representations of the test CMR images fed through the segmentation model encoder.}
\begin{tabular}{|c|c|} \hline
\textbf{Training dataset split} & \textbf{Black vs White}  \\ \hline
100\%/0\%              & 0.950           \\ \hline
75\%/25\%              & 0.950           \\ \hline
50\%/50\%              & 0.949           \\ \hline
25\%/75\%              & 0.951           \\ \hline
0\%/100\%              & 0.949          \\ \hline
\end{tabular}
\label{tab: logistic regression acc}
\end{adjustbox}
\end{table}

The conclusion from this experiment is that race appears to be encoded in the latent representations of the trained segmentation models.

\subsection{Experiment 2: Localisation of source of bias}

The first set of experiments resulted in high accuracy for race classification, suggesting a strong distributional shift. They also suggested that the source of the bias was mainly in the images and that it was being encoded into the segmentation model. Therefore, we next sought to understand which parts of the images were leading to the distributional shift and hence the bias. To visualise the relative importance of the different regions of the image, we used GradCAM \cite{SelvarajuGrad-CAM:Localization} applied to the race classification models.

\begin{figure}[h!]
\includegraphics[width=12cm]{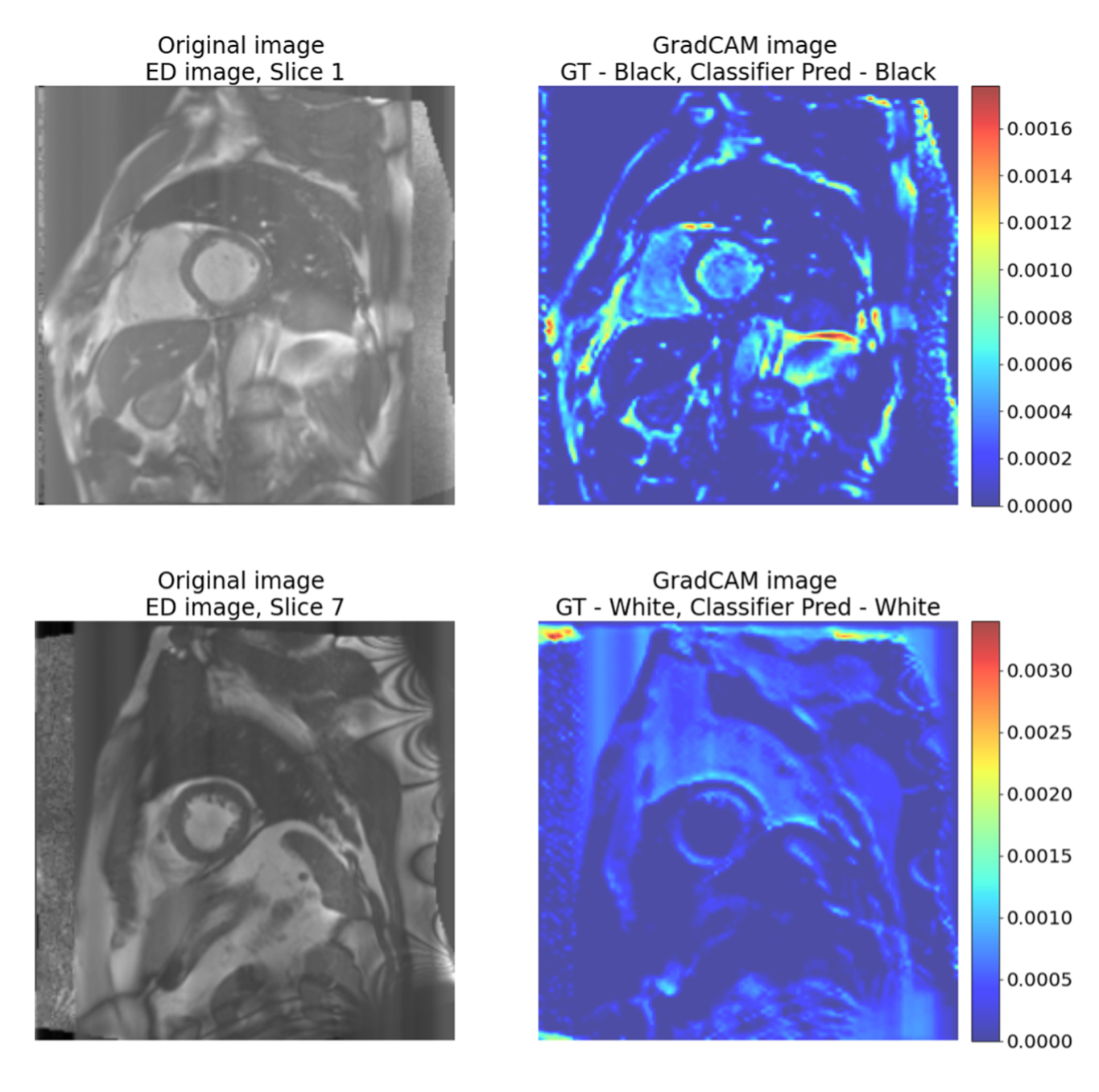}
\centering

\caption{Examples of the normalised CMR images and GradCAM images for the classification model trained on the Im-Im-Im dataset for Black vs White subjects. Higher values (red) correspond to important areas used for race classification; lower values (blue) correspond to less important areas. The top image displays a heatmap where the non-heart regions have higher activations, the bottom image shows a heatmap where artefacts have higher activations.}
\label{fig: gradcam images}
\end{figure}

The results are shown in \cref{fig: gradcam images} with normalised CMR images and GradCAM images. These representative examples show that for both the Black and White subjects, the most attention is being given to non-heart regions such as subcutaneous fat. Further examples can be seen in \cref{fig: gradcam white ex} and \cref{fig: gradcam black ex}.

By visual inspection of all test images, we found that 42\% of the images had the highest activations in non-heart anatomical regions of the body whereas only 6\% had the highest activation in heart regions. The remaining 52\% could be classified as ‘activations due to image artefacts’ (50\%) and ‘other’ where there were no clear activations in any particular area (2\%). These image artefacts become visible after normalising the images which occurs before model training. The artefacts can be caused by interactions between the magnetic field and body tissues during MR image acquisition. For example, ‘ghosting artefacts’ can cause the skin and fat layers to appear as echoes at regular intervals in an image \cite{Alfudhili2016CurrentTricks}. 

\begin{figure}[t]
\includegraphics[width=10cm]{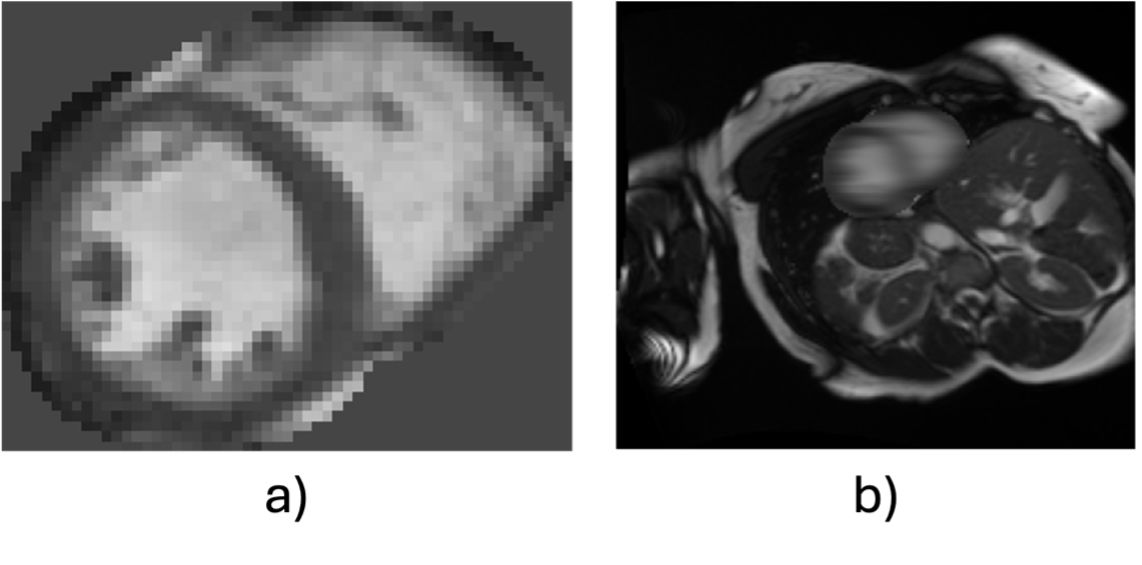}
\centering
\caption{Examples of images including and excluding the heart. a) image cropped around the heart b) image with the heart blurred}
\label{fig: blurred cropped ims}

\end{figure}

Based on these results, we next investigated the impact on race classification performance of using different areas of the images as input. We created two further datasets of images: a dataset including only the heart and a dataset excluding the heart. For the first dataset we cropped the images around the region of the heart using a bounding box based on the ground truth segmentations. All images for a given experiment were cropped to the same size, i.e. the size of the largest heart in the dataset. For the second dataset the heart was blurred out using a Gaussian filter. Examples of the images created in this way are shown in \cref{fig: blurred cropped ims}. We then repeated the race classification experiments using these new images and compared to the performance on uncropped images.

As before, the results were averaged over 10 repeat runs with different random training and validation sets and random seeds for training. The results can be seen in \cref{tab: classification acc cropped}. Cropping the images around the heart regions caused the accuracy to decrease by 0.405 to 0.554. Blurring the images only caused the accuracy to decrease by 0.046 to 0.913. 

\begin{table}
\centering
\begin{adjustbox}{width=\columnwidth,center}

\caption{Classification accuracy for original images, images cropped around the heart and images with the heart blurred. The results show the mean (standard deviation) over 10 repeat runs. The highest result for each measure is shown in bold}
\begin{tabular}{|c|c|c|c|} \hline
\textbf{Image type}             & \textbf{Accuracy}       & \textbf{Sensitivity}             & \textbf{Specificity}    \\ \hline
Im-Im-Im                        & \textbf{0.959 (0.004)~} & \textbf{0.966 (0.013)}~\textbf{} & 0.951 (0.014)~          \\ \hline
Images cropped around the heart & 0.554 (0.028)           & 0.618 (0.061)                    & 0.537 (0.029)           \\ \hline
Images with heart blurred       & 0.913 (0.007)           & 0.884 (0.010)                    & \textbf{0.952 (0.014)} \\ \hline
\end{tabular}
\label{tab: classification acc cropped}
\end{adjustbox}
\end{table}

The conclusion of these experiments is that the main source of the distributional shift for races is outside the heart area. This agrees with the GradCAM experiments which showed higher activation in non-heart regions such as subcutaneous fat and image artefacts.

Based on this conclusion, we next investigated the impact of training segmentation models after cropping out the regions of the images which seemed to be leading to the distributional shift between races, i.e. the areas outside the heart. Specifically, the segmentation experiments performed in \cite{Lee2023AnSegmentation} (using the full CMR images) were repeated using the cropped images. These experiments trained multiple nnU-Net segmentation models using different levels of race imbalance in the training set and evaluated their performance separately for White and Black subjects. The images here were cropped to the same size as the previous classification experiment. The models were trained using the same training parameters as in \cite{Lee2023AnSegmentation}.

\begin{figure}[t]
\includegraphics[scale=0.8]{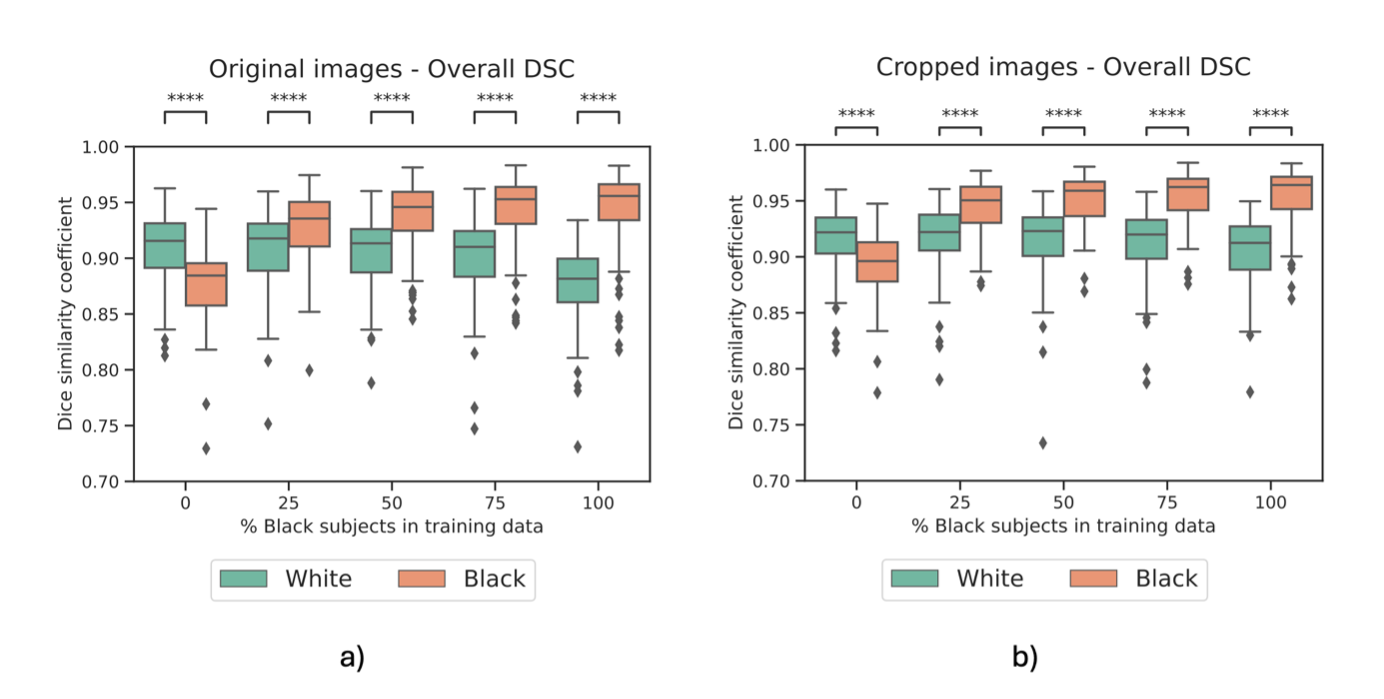}
\centering
\caption{ Overall Dice similarity coefficient (DSC) for segmentation experiments using original (a) and cropped (b) CMR images. Statistical significance was tested using a Mann-Whitney U test and is denoted by **** (p $\leq$ 0.0001), *** (0.001 $<$  p $\leq$ 0.0001), ** (0.01 $<$ p $\leq$ 0.001), * (0.01 $<$  p $\leq$0.05), ns (0.05 $\leq$ p).}
\label{fig: cropped DSC}
\end{figure}

\cref{fig: cropped DSC} shows the results of the segmentation experiments. Comparisons of the predicted end systolic volumes and ejection fraction for the original and cropped images can be seen in \cref{fig: cardiac function measures}. Compared to using the original images, using the cropped images reduced the range of DSCs for both protected groups at each level of training set imbalance. The differences between DSCs of the two protected groups also reduced although the results remain significantly different. Therefore, cropping the images reduced but did not remove the difference in performance between the Black and White subjects. This is consistent with the classification experiments \cref{tab: classification acc} which showed that some distributional shift was present in the heart region.

\subsection{Experiment 3: Are the biases observed in AI CMR segmentation due to confounders?}

Differences in covariates between protected groups may lead to distributional shifts and consequent bias in the AI CMR segmentation models. We investigate whether this is the case by comparing the DSC and the covariates of the subjects such as their weight, height and heart rate. The data is analysed by fitting a linear regression model between covariates and DSC scores.

\cref{tab: confounder table} shows the results of this analysis. The table shows the standardised $\beta$ coefficients and p-values for each covariate. The standardised $\beta$ coefficient shows the relative effect of each covariate on the DSC score, with positive coefficients indicating a positive correlation and negative coefficients indicating a negative correlation. The year of the MRI scan was the most predictive of DSC score in the Black subjects. White subjects have no confounders with a p-value less than 0.05.

\newcolumntype{N}{>{\centering\arraybackslash}m{.5in}}
\newcolumntype{G}{>{\bfseries\centering\arraybackslash}m{2in+6\tabcolsep}}

\begin{table}
\begin{center}
\begin{adjustbox}{width=\columnwidth,center}
\caption{Parameters from a linear regression model fitting DSC scores from a segmentation model trained on
original, uncropped CMR images to covariates for Black and White test subjects. The DSC scores are from an
evenly balanced training dataset. * None of the White subjects had diabetes.}
\begin{tabular}{|c|c|c|c|c|}\hline
\multirow{3}{*}{Covariate}               & \multicolumn{2}{c|}{Black subjects}    & \multicolumn{2}{c|}{White subjects}           \\ \cline{2-5}
                         & Standardised & \multirow{2}{*}{p-value} & Standardised  & \multirow{2}{*}{p-value} \\ 
                         & $\beta$ coefficient &  & $\beta$ coefficient  & \\ \hline
                         
Age                      & 0.060                     & 0.718   & 0.056                      & 0.754   \\ \hline
Height                   & 0.126                     & 0.881   & 0.935                      & 0.227   \\ \hline
Weight                   & 0.240                     & 0.865   & 2.017                      & 0.206   \\ \hline
BMI                      & 0.098                     & 0.940   & 1.832                      & 0.172   \\ \hline
Heart rate               & 0.241                     & 0.063   & 0.284                      & 0.125   \\ \hline
Systolic blood pressure  & 0.253                     & 0.333   & 0.213                      & 0.357   \\ \hline
Diastolic blood pressure & 0.125                     & 0.604   & 0.424                      & 0.138   \\ \hline
LVSV                     & 0.581                     & 0.009   & 0.121                      & 0.713   \\ \hline
LVEF                     & 0.410                     & 0.012   & 0.079                      & 0.734   \\ \hline
LVEDM                    & 0.453                     & 0.057   & 0.141                      & 0.688   \\ \hline
HDL cholesterol          & 0.321                     & 0.015   & 0.002                      & 0.990   \\ \hline
Cholesterol              & 0.015                     & 0.929   & 0.071                      & 0.657   \\ \hline
Diabetes                 & 0.236                     & 0.155   & \multicolumn{2}{c|}{*}                \\ \hline
Hypertension             & 0.032                     & 0.838   & 0.169                      & 0.280   \\ \hline
Hyper cholesterolemia    & 0.175                     & 0.393   & 0.767                      & 0.446   \\ \hline
Smoking                  & 0.1744                    & 0.155   & 0.041                      & 0.776   \\ \hline
MRI year                 & 0.493                     & 0.001   & 0.038                      & 0.808   \\ \hline
\end{tabular}
\label{tab: confounder table}
\end{adjustbox}
\end{center}
\end{table}


\clearpage 
\section{Discussion}
\label{sec:discussion}

In this paper, we have shown that race can be predicted from single SAX CMR images with very high accuracy. However, the accuracy of predicting race from CMR segmentations was noticeably lower, indicating that the distributional shift between White and Black protected groups is mostly in the CMR images as opposed to the manual segmentations. 

The GradCAM images showed that the classification networks had the highest activations in non-heart regions such as subcutaneous fat and image artefacts, a result that was further demonstrated by the classification experiments using a dataset with the heart “removed” from the images. The accuracy here remained high whereas we found low classification accuracy using images cropped tightly around the heart. This suggests that there are fewer race-specific features in the images of the hearts of White and Black subjects and that the distributional shift is mostly in non-heart regions. A similar result was found in \cite{Gichoya2022} where occluding regions identified by saliency maps as important for race classification from chest X-ray images caused the accuracy to decrease.

When looking at segmentation tasks, the high classification accuracy of the logistic regression model showed that subjects’ races were encoded in the latent representations of the CMR images, which makes this encoding a likely cause of the bias in segmentation performance. Cropping the images in a similar fashion to the classification experiments reduced, but did not eliminate, the bias found in the segmentation experiments. We speculate that the remaining bias is due to some anatomical differences in the heart region and the fact that it was not possible to completely crop out non-heart regions in all images because of the variability in heart size and the need to maintain a constant image size for AI model training. 

The covariate analysis indicated that some variables seem to be acting as confounders. LVEDM, LVSV and LVEF were correlated with DSC score for Black subjects but not for White subjects. Black and White subjects are known to have differences in body composition such as fat distribution and bone density \cite{Wagner2000MeasuresReview} as well as differences in cardiac anatomy such as Black subjects having higher left ventricular mass \cite{Nardi2017}. These distributional shifts may be recognisable to a model and may be used for classification tasks and lead to bias in segmentation tasks. MRI year was also a confounder for Black subjects. We further investigated this and found that, by chance, the White subjects selected in our dataset were on average scanned in earlier years than the Black subjects. It is possible that there were differences in image artefacts over time due to small changes in the acquisition protocol, which would be consistent with the GradCAM activations focusing on artefact areas 50\% of the time. Therefore, we reran the segmentation experiments (using the different levels of race imbalance as in Experiment 2) using data which were also matched by MRI year but found no difference in bias characteristics or performance (see \cref{fig: MRI DSC}). Therefore, we conclude that this was a spurious confounding effect caused by random selection of White subjects who were scanned earlier.

As a recommendation for future development of AI CMR segmentation tools, we suggest that training models using images cropped around the heart may be beneficial. However, this does raise the question of how best to crop images in this way at inference time, when ground truth segmentations are obviously not available. Region-of-interest detection methods such as Mask R-CNN \cite{He2017MaskR-CNN} may be useful for this purpose. We also emphasise that such an approach should not be seen as a substitute for more equal representation in CMR datasets. Our experiments have focused on Black and White subjects but previous work \cite{Lee2022ASegmentation} has shown that similar bias effects exist for Asian subjects and by sex \cite{Lee2023AnSegmentation}. Therefore, we argue for greater representation of all protected groups in CMR datasets.

\section{Conclusion}

We have performed a series of experiments to investigate the cause of AI CMR segmentation bias. Our conclusions are (i) the distributional shift between White and Black subjects is mostly, but not entirely, in the images rather than the segmentations, (ii) differences in body fat composition outside of the heart are a likely cause of the distributional shift and hence the bias, (iii) cropping the images around the heart reduces but does not eliminate the bias. Our results will likely be valuable to researchers aiming the train fair AI CMR segmentation models in the future.

\section*{Acknowledgements}
This work was supported by the Engineering \& Physical Sciences Research Council  Doctoral Training Partnership (EPSRC DTP) grant  EP/T517963/1. This research has been conducted using the UK Biobank Resource under Application Number 17806.


\bibliography{refs}{}
\bibliographystyle{ieeetr}

\newpage
\clearpage
\section*{Supplementary Material}
\label{sec:supplementary}

\renewcommand{\thetable}{S\arabic{table}}
\renewcommand{\thefigure}{S\arabic{figure}}

\setcounter{figure}{0}
\setcounter{table}{0}

\subsection{Experiment 1: source of the bias}

\begin{figure}
    \includegraphics[width=13cm]{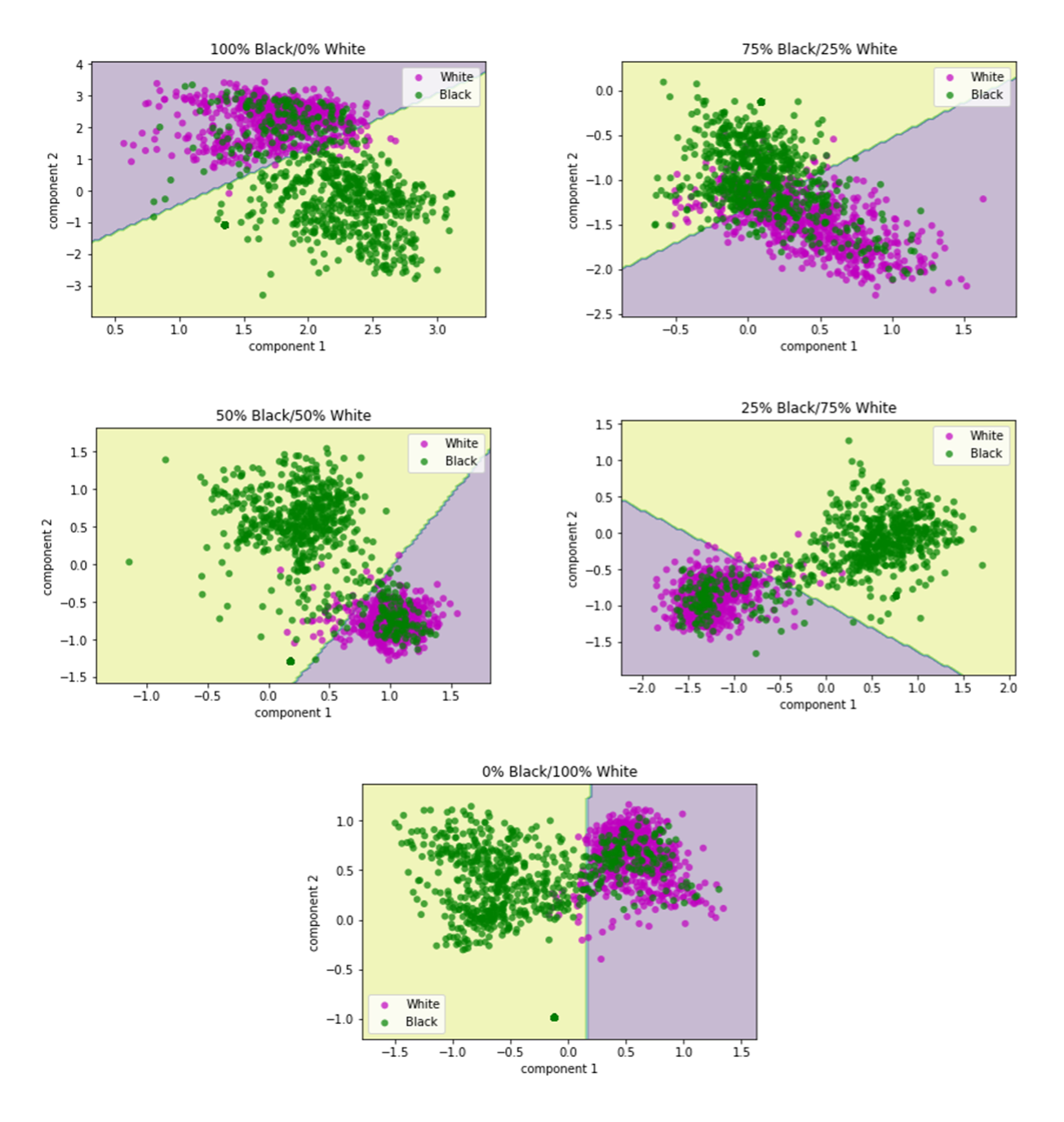}
    \centering
    \caption{Component 1 and 2 of PCA on latent space representations of CMR images from nnU-Net.}
    \label{fig: latent space classification}

    \end{figure}

\begin{figure}[t]
    \includegraphics[width=11cm]{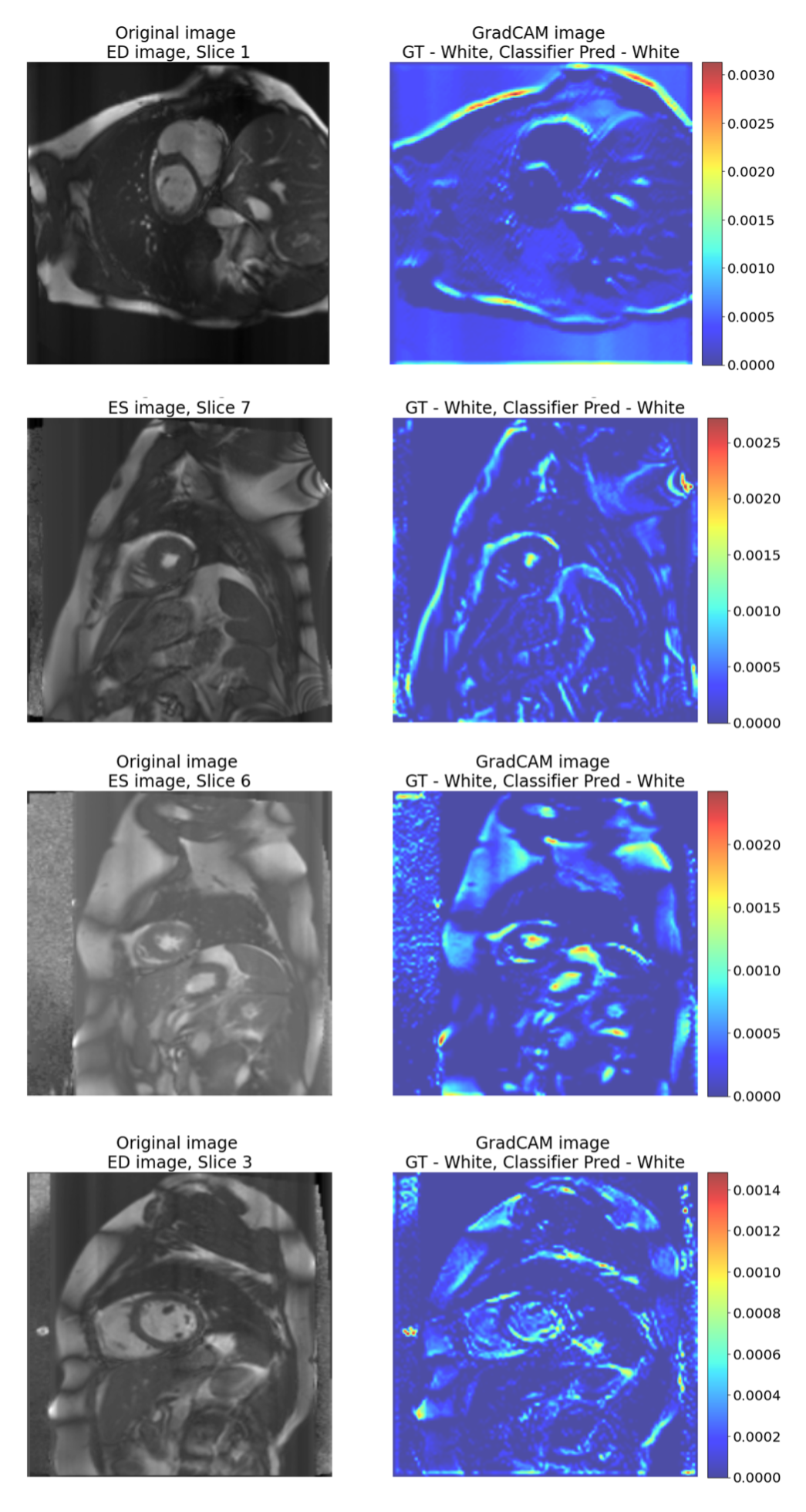}
    \centering
    \caption{Examples of normalised CMR images and GradCAM heatmaps for Im-Im-Im dataset for the White subjects in race classification experiments}
    \label{fig: gradcam white ex}

    \end{figure}

\begin{figure}[t]
    \includegraphics[width=11cm]{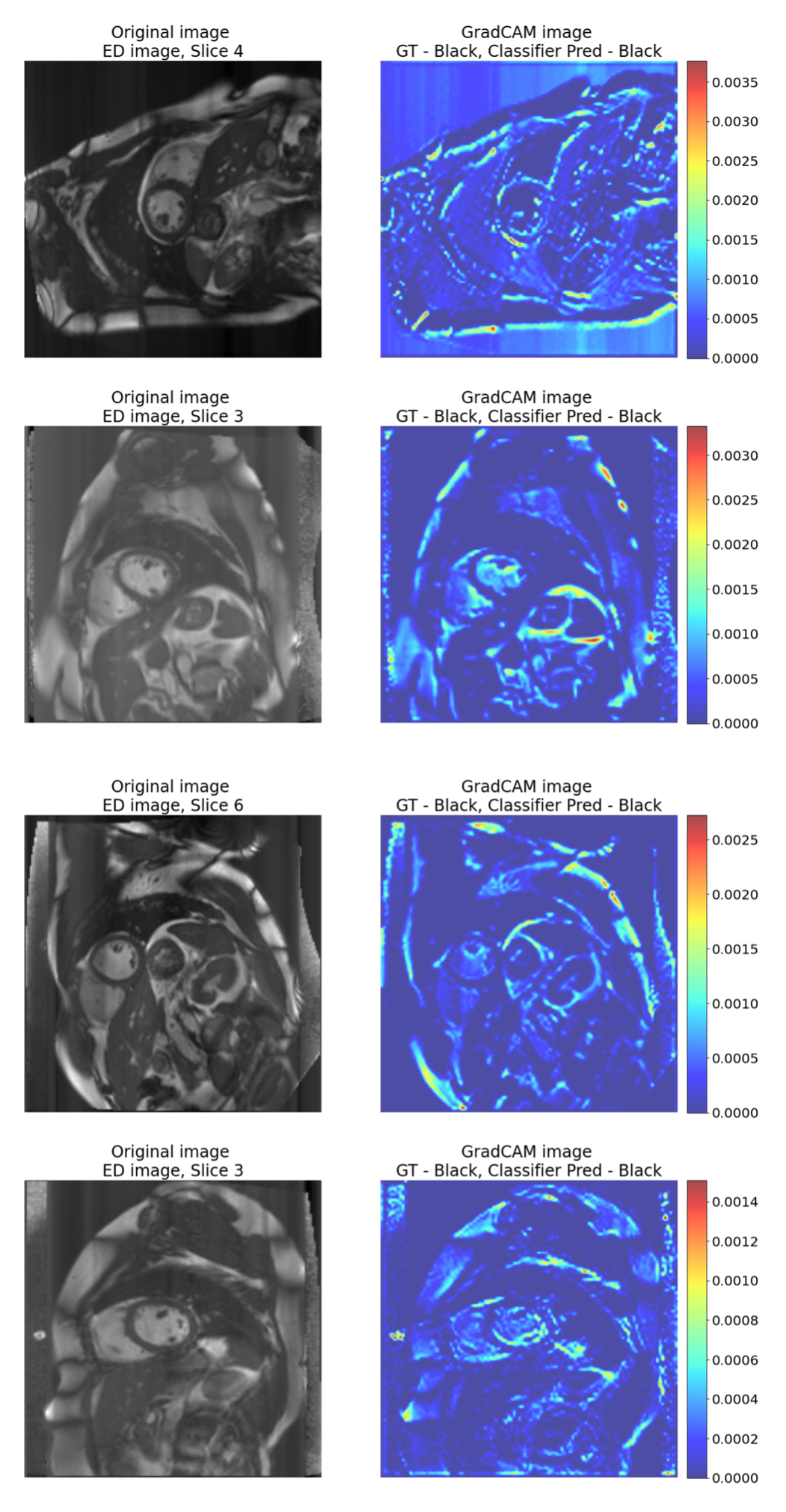}
    \centering
    \caption{Examples of normalised CMR images and GradCAM heatmaps for Im-Im-Im dataset for the Black subjects in race classification experiments}
    \label{fig: gradcam black ex}
    \end{figure}

\begin{enumerate}
    \item For protected attribute classification, all datasets were trained using a model which was pre-trained on images from the ImageNet1k dataset, apart from the Seg-Seg-Seg dataset which was trained from scratch using randomised weights. 
    \item A visual representation of the decision boundary used for the classification of latent space representations of images can be seen in \cref{fig: latent space classification}.
\suspend{enumerate}

\subsection{Experiment 2: localisation of the source of the bias}
\resume{enumerate}

    \item Before plotting, all GradCAM heatmaps were smoothed using a Gaussian blur with kernel size (3,3) and standard deviation chosen from a uniform distribution between 1 and 2 chosen by visual inspection. Further examples of GradCAM images can be seen in \cref{fig: gradcam white ex} and \cref{fig: gradcam black ex} using the Im-Im-Im dataset.
\end{enumerate}

      \begin{figure}[t]
    \includegraphics[width=12cm]{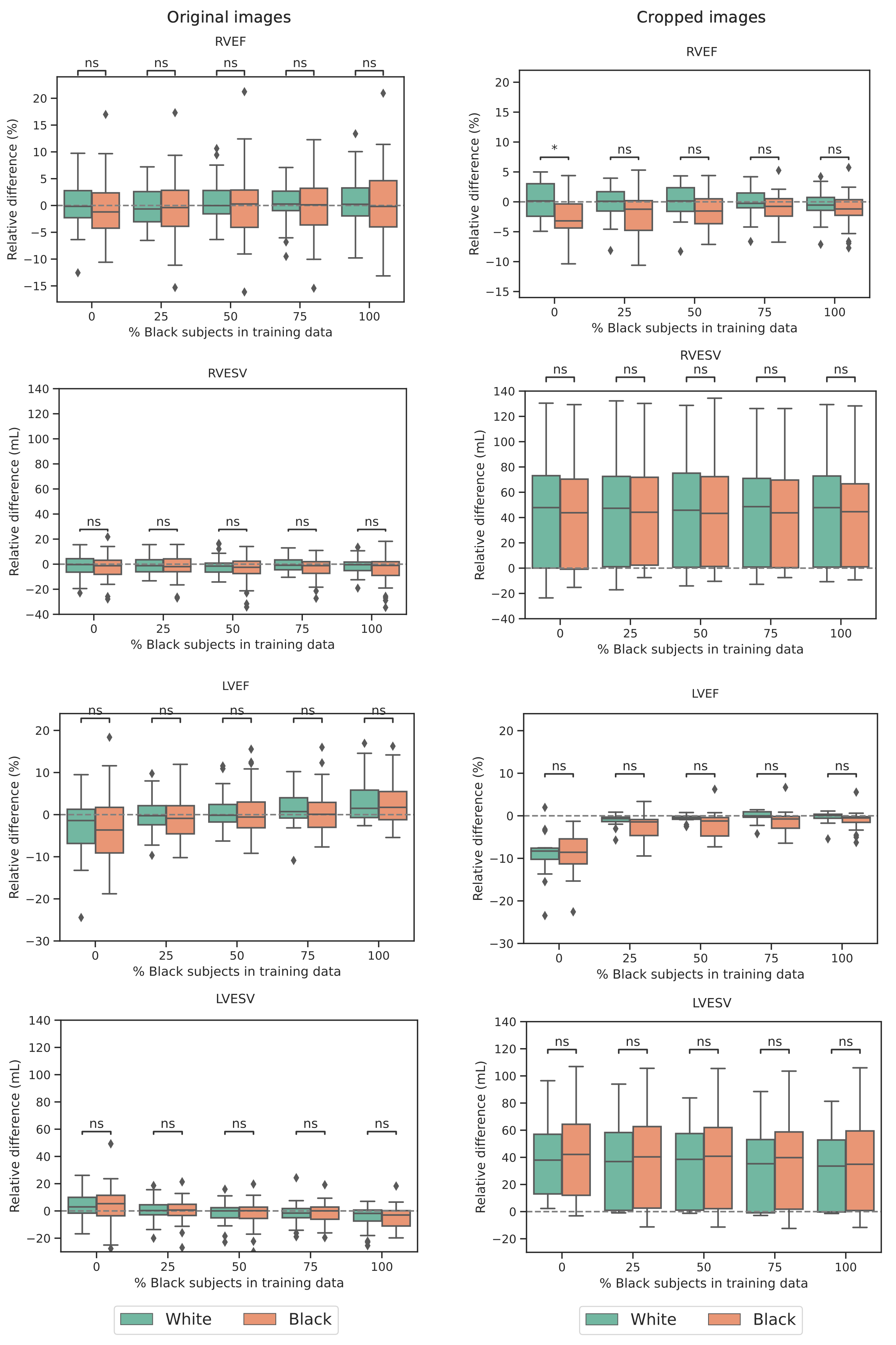}
    \centering
    \caption{Comparison of the prediction of right ventricular ejection fraction (RVEF), right ventricular end systolic volume (RVESV), left ventricular ejection fraction (LVEF) and left ventricular end systolic volume for the nnU-Net model using original images (left column) and cropped images (right column). Statistical significance was tested using a Mann-Whitney U test and is denoted by **** (p $\leq$ 0.0001), *** (0.001 $<$  p $\leq$ 0.0001), ** (0.01 $<$  p $\leq$ 0.001), * (0.01 $<$  p $\leq$ 0.05), ns (0.05 $\leq$ p). }
    \label{fig: cardiac function measures}

    \end{figure}

    \begin{figure}[t]
    \includegraphics[width=11cm]{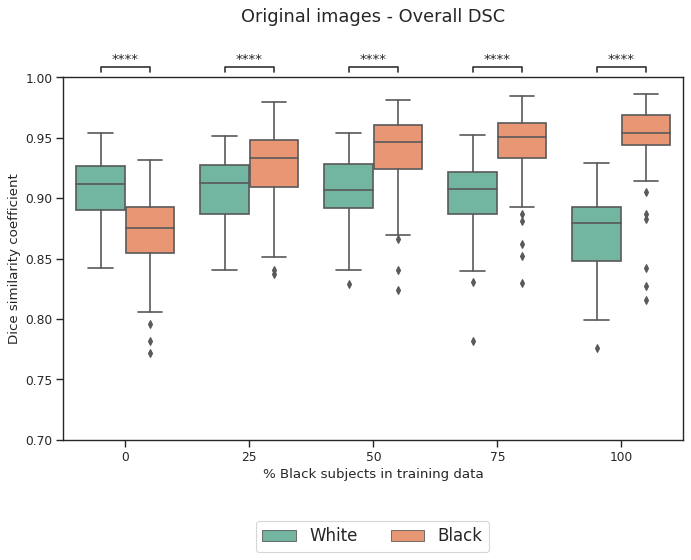}
    \centering
    \caption{Overall Dice similarity coefficient (DSC) for segmentation experiment using original CMR images. The subjects are controlled by age and MRI year. Statistical significance was found using a Mann-Whitney U test and is denoted by **** (p $\leq$ 0.0001), *** (0.001 $<$  p $\leq$ 0.0001), ** (0.01 $<$  p $\leq$ 0.001), * (0.01 $<$  p $\leq$ 0.05), ns (0.05 $\leq$ p).}
        \label{fig: MRI DSC}

    \end{figure}

\end{document}